# Personal Mobile Malware Guard PMMG: a mobile malware detection technique based on user's preferences

"The author declares that there is no conflict of interest regarding the publication of this paper."
**Belal Amro**
College of Information Technology, Hebron University Hebron, West Bank, Palestine

**Abstract**
Mobile malware has increased rapidly last 10 years. This rapid increase is due to the rapid enhancement of mobile technology and their power to do most work for their users. Since mobile devices are personal devices, then a special action must be taken towards preserving privacy and security of the mobile data. Malware refers to all types of software applications with malicious behavior. In this paper, we propose a malware detection technique called Personal Mobile Malware Guard – PMMG- that classifies malwares based on the mobile user feedback. PMMG controls permissions of different applications and their behavior according to the user needs. These preferences are built incrementally on a personal basis according to the feedback of the user. Performance analysis showed that it is theoretically feasible to build PMMG tool and use it on mobile devices.

## 1. Introduction

According to Oxford dictionary, malware is defined as "Software which is specifically designed to disrupt, damage, or gain authorized access to a computer system. ". this category of software includes all types of software with malicious intent like Trojans, Viruses, Worms, … etc. The most common malware programs are viruses. A virus is a self-replicating code that can infect programs by modifying them or their environment, it is able to spread rapidly via email or network propagation. A worm is an independent program that copy itself and spreads over the network, the new copies are fully independent and can spread by their own [1].

A Trojan is a software that appears to the user to be benign application however, it performs malicious acts in the background [6]. Trojan are used to help attacking a system by performing acts that might compromise security of the system and hence enables hacking it easily. Ransomware is another type of malware that prevents the users from accessing their data by locking the device or encrypting the data files, until ransom amount is paid [10]

A famous malware is called a "bot" which is a type of malware that enables an attacker to take control over an affected Mobile device, it is also known as "Web robots", they are part of a network of infected machines, known as a "botnet", which is typically made up of all victim mobile devices across the globe[11]. Spywares are simply spying software. They run unnoticed in the background while they collect information, or give remote access to their authors [12] ,[13] [16].

The number of mobile malwares is increasing dramatically last two years. According to Macafe LABs [28], the number of malwares exceeded 16,000,000 in first quarter of 2017. By looking at the global mobile malware infection rate reported by Macafe LABs 2017, Figure 1 shows a significant increase in the infection rate for the first quarter of the year 2017.

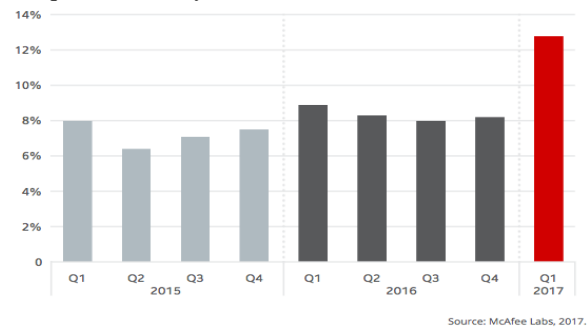

Figure 1: global mobile malware infection rates

Kaspersky Labs [32] reported the distribution of new mobile malware in the years 2015 and 2016 as shown in Figure 2:

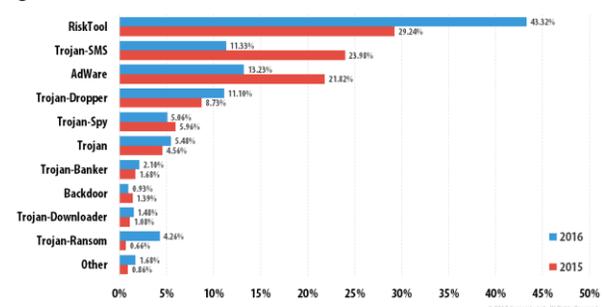

Figure 2: distribution of mobile malware

These types of malware are harmful to systems and hence must be detected and removed to make sure that the system functions well. The rest of the paper is organized as follows: A literature review of current malware detection techniques is proposed. A description of how PMMG technique works is then provided. Analysis of the technique is carried out. And at last a summary is provided.

## 2. Literature review

To better detect malwares, we have to both understand their behavior and how do they spread, in this section, we provide a brief summary of how do malwares spread. We also provide the state-of-the-art mobile malware detection technique.

2.1 Malware spreading techniques

To mitigate malware attacks, we should be aware of malware spreading techniques. In this section, we categorize malware spreading techniques as follows:

2.1.1 Repackaging

Malware authors repackage popular mobile applications in official market, and distribute them on other less monitored third party markets. Repackaging includes the disassembling of the popular benign apps, and appending the malicious content and reassembling. This is done by reverse-engineering tools. TrendMicro report have shown that 77% of the top 50 free apps available in Google Play are repackaged [14].

2.1.2 Drive By Download

Drive by Download refers to an unintentional download of malware in the background. It Occurs when a user visit a website that contains malicious content and downloads malware into the device. Android/NotCompatible [15] is the most popular mobile malware of this category.

2.1.3. Dynamic Payloads

Uses dynamic payload to download an embedded encrypted source in an application. After installation, the app decrypts the encrypted malicious payload and executes the malicious code [16].

2.1.4. Stealth Malware Techniques

Stealth Malware Technique refers to an exploit of hardware vulnerabilities to obfuscate the malicious code to easily bypass the antimalware. Different stealth techniques such as key permutation, dynamic loading, native code execution, code encryption, and java reflection are used to attack the victim's device [16].

2.2 Malware detection techniques:

In this section, we analyze the state-of-the-art malware detection techniques for mobile devices. According to [30], mobile malware detection techniques are categorized into two categories according to the basis they rely on when detecting for malwares. The categories are statics techniques and dynamic techniques

2.2.1 Static techniques:

Static techniques rely on the source code of an application to classify it accordingly without having the application executed. These techniques are classified into one of the following classes according to the basis they rely on for analyzing the source code. Table 1 summarizes static techniques

Table 1: summary of mobile malware static detection techniques

| Technique | How does it work | Advantages | Disadvantages |
|---|---|---|---|
| Signature Based Approach [18]. | This method extracts the semantic patterns and creates a unique signature | It is a very fast method for detecting malware | IT can only identify the existing malwares and fails against the unseen variants of malwares |
| Permission Based Analysis [19]. | Analyzes permissions required by applications and detect abnormal requirements | fast in application scanning and identifying malware | Permission based methods require second pass to provide efficient malware detection. |
| Virtual machine analysis [20]. | tests the app behavior and analyses control and data flow which in sake of detecting dangerous functionalities | Tests the byte code of an application and track sensitive API calls | Analysis is performed at instruction level and consumes more power and storage space. |

Table 2: Dynamic mobile malware detection techniques

| Technique | How does it work | Advantages | Disadvantages |
|---|---|---|---|
| Anomaly based [21],[22],[23],[24] | based on watching the behavior of the device by keeping track of different parameters and the status of the components of the device | It engages different parameters and hence have a clear image of the system | The larger the parameters engaged the more the calculation required. |
| Taint analysis[25] | tracks multiple sources of sensitive data and identifies the data leakage in mobile apps | efficient tracking of sensitive data | does not perform control flow tracking. |
| Emulation based [26] | dynamically analyze apps based on Virtual Machine Introspection | It monitors the whole system by being out of execution environment | cannot detect new malwares |

2.2.2 Dynamic techniques:

In dynamic analysis, an application is examined during execution and then classified according to one of the following techniques. The classification is done according to the behavior of the detection mechanism. Table 2 summarizes these techniques.

As shown in this Section, malware detection techniques have limitations and do not give very accurate detection. The idea is to engage the user's preferences in malware detection by having their feedback from interactions they provide when using the malware detection tool. In the next section, we will provide details about our proposed Personal Mobile Malware Guard – PMMG.

## 3. PMMG System

Our proposed malware detection technique depends on user's preferences for application permissions. The technique is called Personal Mobile Malware Guard, shortly PMMG. PMMG works in between the operating system and mobile applications. It interacts with the mobile user for granting permission to the application. In case the user denies the permission, PMMG refuses the permission, and incase a program will terminate if permission is denied, then PMMG provides a virtual resource to the application so that it will not have actual access to the resource and will continue working with that virtual resource. In this section, we will detail the components of PMMG and the work flow as well.

As shown in Figure 3, PMMG consists of four modules that interact with the mobile user and application. These modules are PMMG interface, Permit Granter, Virtual profile, and Rule Base.

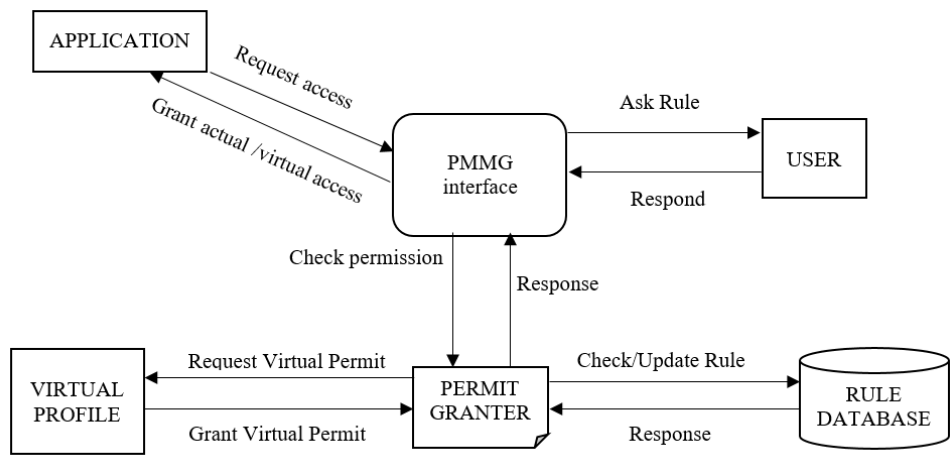

Figure 3: components of PMMG system

3.1 PMMG Components:

The components of PMMG shown in Figure 3 will be described in this subsection, the workflow will be detailed in the later subsection.

3.1.1 PMMG interface:

The PMMG interface is the module that interacts with user and the mobile application. The role of this module is to provide an interface to both mobile application and mobile user. The mobile application will have to ask PMMG interface for a permission. The mobile interface will interact with user and with the other components of PMMG system and respond to the application by granting access to the resource or blocking the access and pretend to make it available to the application by a virtual access profile (will be detailed later).

### 3.1.2 Permit Granter:

The permit granter is responsible for issuing the decision about the permission required according to user's feedback and the rule database. It forms the gateway of the PMMG interface with both Virtual Profile and Rule database. The aim of this architecture is to hide the Rule database and Virtual profile from the PMMG interface and hence from the mobile application. The granter will check the rule database and respond to the PMMG interface by a suitable response accordingly.

### 3.1.3 Rule Database

The Rule Database contains the permission rules granted to applications. Each application requires a permission and granted that permission is stored in the database. Applications with denied permissions are also stored. The rule database returns the permission status of that application and responds to the PMMG interface. Permission state might be actual or virtual. Actual permissions are those real permissions granted to application while virtual permission are fake permission provided to application so that they can proceed working with the least required permission.

### 3.1.4 Virtual profile

The Virtual Profile module is responsible for running a virtual process simulating interface of the actual resource required. For example, if a program requires a permission to the microphone and the user denied that access, this module will run a process simulating the work of the microphone without having to run the actual microphone. For Virtual Profile to work, it requires building virtual profiles for each sensitive resource in the system including microphone and camera. When a user refuses a permission to the program and still need that program to work, then the virtual profiler will deceive the program by launching the virtual profile instead of actual profile.

For a virtual profiler to work well, it has to support virtual drivers for sensitive hardware resources such as camera and mic. There are many mobile applications that support this and can easily be found at google play and app store. Besides, a virtual profiler has to feed fake data for sensitive components like contacts, logs, messages … etc. As for hardware, virtual drivers, there are many application for these sensitive applications on google play and app store.

### 3.2 PMMG Work Flow

Base on Figure 3, the work flow of PMMG consists of interactions among PMMG interface, the mobile user, and the mobile application. Before proceeding with the work flow, the assumptions of the underlying system must be first addressed. In this subsection, we will provide the assumption of the underlying mobile operating environment and then we will detail the work flow.

### 3.2.1 Environment assumptions

With PMMG, we assume that the underlying mobile operating system delegates the management of permissions to PMMG by allowing it to grant access, deny access, or grant access virtually. Grant access gives the application full access to the required resource, while deny access prohibit access to the resource. Grant access virtually denies actual access to the resources and initiates a virtual resource to deceive the application of granting it full access.

### 3.2.2 Work flow

To have PMMG working properly, it has to be set up and initialized properly. Therefore, there are two phases for PMMG to work, set up phase and running phase. In this subsection, we will detail both phases.

#### 3.2.2.1 Setup phase

In the setup phase, PMMG will be granted full access to the permission file of the mobile operating system. PMMG should build virtual profiles for all sensitive mobile resources. These virtual profiles are built upon the underlying operating system and must have the same interface of the resource required. The application will interact with the virtual profile as if it is interacting with the actual resource and the virtual profile should respond with virtual response similar to the actual profile response. For example, if the resource is the camera, the virtual profile should have the same interface of the camera and respond with similar, but fake, images to the application. Since the camera is denied by user, then the application will continue working with using virtual camera instead of actual one, hence user's privacy will be maintained. This is applied to all other resources like microphone, contacts, messages, Wi-Fi … etc.

#### 3.2.2.2 Running Phase

During running phase, PMMG is expected to guard the mobile resources according to user's preferences. The running phase is described below:

1) When a mobile application is newly installed, all required permission are transferred to PMMG interface which in turns invoke the user for the permissions.
2) According to the user's responses, permissions are either:
    a. Granted
    b. Denied
    c. Granted virtually – when the program does not work without a permission to this resource.
3) The user's preferences rules are then transferred to Permit Granter who in turns stores these rules in Rule Database.
4) When a mobile application is opened, the permissions are transferred to PMMG interface for decisions.
5) PMMG interface then asks Permit Granter to check Rule Database for those permissions.
    a. If permission status is grant, then PMMG interface will grant that permission to application.
    b. If permission status is denying, then PMMG interface will deny that permission to application.
    c. If permission status is grant virtually, then PMMG interface will ask Virtual Profiler for a virtual grant interface.
6) If permission status is not available in Rule Database, then PMMG interface will ask the user for the status of that permission. And the steps 2 and 3 are performed.
7) PMMG interface enables users to modify the rules according to their preferences, once modified, these rules will be applied.

Setup phase and running phase form the basis of our PMMG algorithm. In the next section, we will analyze the performance of PMMG according to these two stages.

## 4. Performance analysis

To analyze the performance of PMMG, we should take into account the cost of setup phase and running phase. The setup phase is performed one time when PMMG is installed, and hence the cost of this phase is constant. The important analysis is the runtime analysis i.e. performance of PMMG during running.

According to [28], the average number of mobile applications used by users in 9 daily and 30 monthly. This means that average users will have less than one new application per day. Based on these statistics, and taking into consideration the status of an application, we can calculate the performance of PMMG according to Table 3.

Table 3: calculating performance of PMMG

| status | Required steps |
|---|---|
| Newly installed program | User interaction (*UI*) Permit Granter (*PG*) Database access (*DBA*) Virtual Profile (optional) (*VP*) |
| Previously installed program | Permit Granter (*PG*) Database access (*DBA*) Virtual Profile (optional) (*VP*) |

According to Table 1, the following formulas calculate the performance for the newly installed program and the previously installed ones.

$$new_{app} = UI + PG + DBA + VP \dots equation\ 1$$
$$old_{app} = PG + DBA + VP \dots equation\ 2$$

The time required for performing *UI*, *PG* and *DBA* is constant and mainly depends on the mobile device specifications. However, the time required for V*P* varies according to the type of the permission required and the time *VP* will run. *VP* will launch a virtual profile that will run instead of the actual resource the whole period required by the application program to run that resource. Besides, t *VP* will run different profiles according to the required resource, hence, different profiles require different running times.

Generally, as the number of applications increases, then the total required time for PMMG increases. Assuming that the number of application for a particular mobile user is *n*, then we have:

$$PMMG_{Daily} = new_{app} + n * old_{app} \dots equation\ 3$$

Since we have 9 applications per user daily, and we have less than one new application daily, then the daily performance will be:

$$PMMG_{Daily} = new_{app} + 9 * old_{app}$$

According to [29], the average user spends about 3 hours on mobile device daily, and knowing that 9 applications are used daily, then, each application on average is run about 20 minutes which we will call *appTime*. Assuming that half of the applications will require *VP* to run, then *VP* time can be calculated as follows:

$$VP = \frac{n}{2} * appTime \dots equation\ 4$$

According to equation 3 and equation 4, we have:
$$PMMG_{Daily} = UI + PG + DBA + VP + (n) * (PG + DBA + VP) \dots equation\ 5$$

Simplifying equation 5 results in:
$$PMMG_{Daily} = UI + (n + 1) * (PG + DBA + VP)$$

And substituting VP will result in:
$$PMMG_{Daily} = UI + (n + 1) * (PG + DBA + \frac{n}{2} * appTime) \dots equation\ 6$$

According to equation 6, the running time of PMMG is quadratic with respect to the number of running applications $n$, however, $n$ is small (on average 9 for the year 2017) and hence the time will not grow large. This makes our PMMG theoretically feasible in terms of performance and this small lack of performance is justified for the sake of better and controlled privacy.

## 5. Summary

In this paper, we proposed a novel mobile malware detection technique called PMMG. PMMG relies on user's preferences to manage mobile application permission in a way to better enhance mobile user's privacy. These preferences are built incrementally on a personal basis according to the feedback of the user. Detailed description of the components and workflow of PMMG technique besides performance analysis showed that applying this technique to detects and block malware access to sensitive mobile resources is feasible, but slightly reduces performance. This small reduction in performance is really justified to increase the privacy and security level of the mobile device and provide a better privacy management to the user. As a future work, a tool based on PMMG will be built and tested. Practical performance results and any other modifications will be proposed in a future work.

## References:

[1] J. Horton and J. Seberry , " Computer Viruses An Introduction " , proceeding of the 20th Australian Computer Science Conference, 1997
[2] web site http://www.gartner.com/newsroom/id/3609817 accessed 29/9/2017
[3] Aijaz Ahmad Sheikh et. al. , Smartphone: Android Vs IOS , The SIJ Transactions on Computer Science Engineering & its Applications (CSEA), September-October 2013
[4] Web site https://developer.apple.com/library/content/documentation/Miscellaneous/Conceptual/iPhoneOSTechOverview/CoreOSLayer/CoreOSLayer.html#//apple_ref/doc/uid/TP40007898-CH11-SW1 last accessed 29/9/2017
[5] Thomas L. Rakestraw et. al., The mobile apps industry: A case study , Journal of Business Cases and Applications, 2013.
[6] "Android and Security - Official Google Mobile Blog." [Online]. Available: https://www.blog.google/topics/safety-security/shielding-you-potentially-harmful-applications/html. [Accessed: 28-sep-2017].
[7] R. Raveendranath, V. Rajamani, A. J. Babu, and S. K. Datta, "Android malware attacks and countermeasures: Current and future directions," 2014 Int. Conf. Control. Instrumentation, Commun. Comput. Technol., pp. 137–143, 2014.
[8] "root exploits." [Online]. Available: http://www.selinuxproject.org/~jmorris/lss2011_slides/caseforseandroid. pdf. [Accessed: 15-Dec-2015].
[9] Y. Zhou, Z. Wang, W. Zhou, and X. Jiang, "Hey, You, Get Off of My Market: Detecting Malicious Apps in Official and Alternative Android Markets," Proc. 19th Annu. Netw. Distrib. Syst. Secur. Symp., no. 2, pp. 5–8, 2012.
[10] "Android.Fakedefender.B | Symantec." [Online]. Available: https://www.symantec.com/security_response/writeup.jsp?docid=2013- 091013-3953-99. [Accessed: 15-Dec-2015].
[11] Y. Zhou and X. Jiang, "Dissecting Android Malware: Characterization and Evolution," 2012 IEEE Symp. Secur. Priv., no. 4, pp. 95–109, 2012.
[12] Y. Zhou and X. Jiang, "Dissecting Android Malware: Characterization and Evolution," 2012 IEEE Symp. Secur. Priv., no. 4, pp. 95–109, 2012.
[13] C. a Castillo, "Android Malware Past , Present , and Future," McAfee White Pap. Mob. Secur. Work. Gr., pp. 1–28, 2011
[14] "A Look at Repackaged Apps and their Effect on the Mobile Threat Landscape." [Online]. Available: http://blog.trendmicro.com/trendlabs- security-intelligence/a-look-into-repackaged-apps-and-its-role-in-the- mobile-threat-landscape/. [Accessed: 15-Dec-2015].
[15] "NotCompatible Android Trojan: What You Need to Know | PCWorld." [Online]. Available: http://www.pcworld.com/article/254918/notcompatible_android_trojan_ what_you_need_to_know.html. [Accessed: 15-Dec-2015].
[16] New Threats and Countermeasures in Digital Crime and Cyber Terrorism. IGI Global, 2015.
[17] "the apple threat landscape" Symantec, [online]. Available: http://www.symantec.com/content/en/us/enterprise/media/security_response/whitepapers/apple-threat-landscape.pdf. [accessed: 29-sep -2017]
[18] A. Aiken, "Apposcopy : Semantics-Based Detection of Android Malware Through Static Analysis," Fse 2014, pp. 576–587, 2014.
[19] Android Permissions Demystified." [Online]. Available: https://www.truststc.org/pubs/848.html. [Accessed: 06-Nov-2015].
[20] Y. Aafer, W. Du, and H. Yin, "DroidAPIMiner: Mining API-Level Features for Robust Malware Detection in Android," Secur. Priv. Commun. Networks, vol. 127, pp. 86–103, 2013.
[21] A. Shabtai, U. Kanonov, Y. Elovici, C. Glezer, and Y. Weiss, ""Andromaly": a behavioral malware detection framework for android devices," J. Intell. Inf. Syst., vol. 38, no. 1, pp. 161–190, 2012
[22] "strace download | SourceForge.net." [Online]. Available: http://sourceforge.net/projects/strace/. [Accessed: 22-Dec-2015].
[23] M. Zhao, F. Ge, T. Zhang, and Z. Yuan, "AntiMalDroid: An efficient SVM-based malware detection framework for android," Commun. Comput. Inf. Sci., vol. 243 CCIS, pp. 158–166, 2011.
[24] Dimitrios Damopoulos et.al. , The Best of Both Worlds. A Framework for the Synergistic Operation of Host and Cloud Anomaly-based IDS for Smartphones, EuroSec'14, April 13 - 16, 2014
[25] W. Enck, P. Gilbert, B.-G. Chun, L. P. Cox, J. Jung, P. McDaniel, and A. N. Sheth, "TaintDroid: An Information-Flow Tracking System for Realtime Privacy Monitoring on Smartphones," Osdi "10, vol. 49, pp. 1– 6, 2010.

[26] L. Yan and H. Yin, "Droidscope: seamlessly reconstructing the os and dalvik semantic views for dynamic android malware analysis," Proc. 21st USENIX Secur. Symp., p. 29, 2012.
[27] T. Bläsing, L. Batyuk, A. D. Schmidt, S. A. Camtepe, and S. Albayrak, "An android application sandbox system for suspicious software detection," Proc. 5th IEEE Int. Conf. Malicious Unwanted Software, Malware 2010, pp. 55–62, 2010.
[28] Website "https://techcrunch.com/2017/05/04/report-smartphone-owners-are-using-9-apps-per-day-30-per-month ", last accessed 03/01/2018
[29] Website "https://www.comscore.com/Insights/Blog/Mobile-Matures-as-the-Cross-Platform-Era-Emerges ", last accessed 03/01/2018.
[30] Belal Amro, "Malware Detection Techniques for Mobile Devices ", International Journal of Mobile Network Communications & Telematics ( IJMNCT), December 2017.

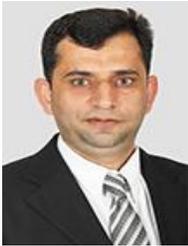

**Belal Amro** is an assistant professor at Faculty of Information Technology in Hebron University - Palestine, where he has been working since 2003. From 2003 to 2004, he was a research assistant at Hebron University. From 2005 to 2007, he was an instructor in the Computer Science Department at Hebron University after having his MSc. degree in complexity and its interdisciplinary applications form Pavia- Italy. During 2008-2011 he received an ERASMUS PhD grant at Sabanci University-Turkey. From 2011-2012 he worked as research assistant at Sabanci University. In 2012, Mr Amro has received his PhD in Computer Science and Engineering From Sabanci University- Istanbul, turkey. From 2012 till now, Mr Amro has been working as an assistant professor at Faculty of Information Technology – Hebron University. From 2014 to 2017 Mr Amro has worked as chairman of Computer Science Department at Hebron University. Mr Amro has served as technical program committee member for different international conferences and journals, and reviewed more than 50 papers in the field of information technology including privacy and security. Currently, Mr Amro is conducting research in network security, wireless security, privacy preserving data mining techniques and has published more than 10 papers in international journals and conferences in the field of computer security and privacy.